\documentclass[12pt]{article}

\def\rv{{\bf r}}
\def\xv{{\bf x}}

\usepackage{amsmath}
\usepackage[super,comma,sort&compress]{natbib}
\usepackage{graphicx}
\usepackage[width=6.75in, height=8.5in, head=0.0in, foot=0.5in, headsep=0.0in]{geometry}
\usepackage[version=3]{mhchem}

\begin{document}
\title{Dispersion with Fixed Diagonal Matrices: Exchange energy correction and an assessment of the Becke--Roussel exchange hole}

\author{\textbf{Timo Weckman}\\
Department of Chemistry, Nanoscience Center,\\
University of Jyväskylä, 40014 Jyväskylä, Finland}

\maketitle

\begin{abstract}
An exchange-correction to the Fixed Diagonal Matrices (FDM) method is introduced to improve accuracy when employing a single reference wavefunction. Also, the performance of the Becke--Roussel exchange-hole for approximating the pair density mediated integrals is explored. With the exchange-correction, the FDM procedure yields dispersion coefficients for closed-shell atoms on par with highly correlated methods when using Hartree--Fock or Kohn--Sham pair density. Conversely, the Becke--Roussel exchange-hole results in an overestimation of the dispersion coefficients for closed-shell atoms. In general, the FDM method fails in underestimating the dispersion coefficients of open-shell atoms and ions.
\end{abstract}

\section{Introduction}

The London dispersion interaction is a weak attractive interaction and essential for chemistry as we know it. This is due to the slow decay of the dispersion interaction: molecules can be weakly bound to each other over distances where the formation of  covalent bonds is not possible. Hence, dispersion interaction plays a vital role in e.g. solid-state physics, protein folding and the structure of DNA.

Dispersion interaction is caused by the quantum-fluctuations in the electron density. While this fluctuation is included in different post-Hartree--Fock methods, it is not properly captured by the present day density functional theory with approximate density functionals and, therefore, different approaches have been proposed for including this interaction in the calculations. The most often employed and computationally most efficient approach is to model the dispersion interaction using a $C_6/R^6$ potential with the dispersion parameters $C_6$ defined for different elements. These parameters may or may not have a density/environment dependency built in.

The Fixed Diagonal Matrices (FDM) method introduces a class of variational wave functions that capture the long-range interactions between two quantum systems without deforming the diagonal of the many-body density matrix of each monomer system\cite{KooGor-JPCL-19}. While this approach can never be exact since the density distortion is explicitly forbidden, it provides a variational expression for the dispersion energy, resulting in an interesting framework for approximating dispersion parameters from the electron density alone. The FDM method performs well for closed shell atoms and molecules\cite{KooWecGor-JCTC-21} when an accurate pair density is used in computing the matrix elements but result in over estimations when an uncorrelated Hartree--Fock/Kohn--Sham wavefunction is employed. 

This paper expands on previous work by introducing an exchange correction to the FDM procedure when employing a single Slater determinant wave function. In addition, we explore the performance the FDM procedure when applying an approximate exchange-hole proposed by Becke and Roussel\cite{BecRou-PRA-89} for computing the pair density dependent terms. These two approaches are benchmarked to a set of closed- and open-shell atoms, ions and molecules, where accurate reference data is available.

\section{Theory}\label{sec:theory}

In the FDM framework\cite{KooGor-JPCL-19}, we consider two systems $A$ and $B$ separated by a large distance $R$ that have isolated ground-state wavefunctions $\Psi^A_0(\underline{\xv}_A)$ and $\Psi^B_0(\underline{\xv}_B)$. Here, $\xv$ denotes the spin-spatial coordinates $(\rv, \sigma)$ and $\underline{\xv}_{A/B}$ denote the whole set of spin-spatial coordinates of electrons in system $A/B$. The dispersion interaction between the two systems is described by the following constrained minimisation problem,
\begin{equation}
\label{eqn:FDM-minization}
E_\text{disp}^{FDM} (R) = \min_{\Psi_R\rightarrow |\Psi_0^A|^2,|\Psi_0^B|^2}\left<\Psi_R\left|\hat{T}+\hat{V}_{ee}^{AB}\right|\Psi_R\right> - T_0^A - T_0^B - U[\rho_0^A,\rho_0^B],
\end{equation}
where $\hat{T}$ is the kinetic energy operator acting on the full set of variables $\underline{\xv}_A, \underline{\xv}_B$, $T_0^{A/B}$ is the ground state kinetic energy expectation value of each subsystem and $U[\rho_0^A,\rho_0^B]$ is the electrostatic repulsion between the ground state densities between the two systems. $\hat{V}_{ee}^{AB}$ is the electron repulsion operator between systems $A$ and $B$, 
\begin{equation}
\hat{V}_{ee}^{AB} = \sum_{i\in A, j\in B} \frac{1}{|\rv_i - \rv_j|}.
\end{equation}
The constraint $\Psi_R\rightarrow |\Psi_0^A|^2,|\Psi_0^B|^2$ means that the minimisation is performed over wavefunctions that leave the diagonal of the many-body density matrix of each monomer unchanged with respect to the ground state value.\cite{KooGor-JPCL-19}

In ref \citenum{KooGor-JPCL-19}, an ansatz wave function was proposed for the minimization, written as
\begin{equation}
\Psi(\underline{\xv_A},\underline{\xv_B}) = \Psi_0^A(\underline{\xv_A})\Psi_0^B(\underline{\xv_B})\sqrt{1+\sum_{i\in A, j\in B} J_R(\rv_i, \rv_j)},
\end{equation}
where the function $J_R$ correlates electrons in $A$ and $B$ but leaves the ground state density of each monomer unchanged,
\begin{align*}
\int \rho_0^{A}(\rv_{i_{A}}) J_R(\rv_{i_A}, \rv_{j_B}) d\rv_{i_A} &= 0 \;\forall\; \rv_{j_B}\\
\int \rho_0^{B}(\rv_{j_{B}}) J_R(\rv_{i_A}, \rv_{j_B}) d\rv_{j_B} &= 0 \;\forall\; \rv_{i_A}.
\end{align*}
Due to this constraint, the external potential and the electron--electron interactions within each monomer cancel out in the interaction energy. 
The function $J_R$ can be expanded as a series of one-body functions $b(\rv)$, dubbed dispersals\cite{KooGor-FD-20}
\begin{equation}
\label{eqn:J-expansion}
J_R(\rv, \rv') = \sum_{ij} c_{ij, R} b_i^A(\rv)b_j^B(\rv'),
\end{equation}
where $c_{ij, R}$ are parameters to be determined variationally. If we perform a multipolar expansion of the monomer--monomer interaction, we can expand the coefficients $c_{ij}$ in a series of inverse powers of $R$ as, 
\begin{equation}
c_{ij,R} = c_{ij}^{(3)}R^{-3} + c_{ij}^{(4)}R^{-4} + c_{ij}^{(5)}R^{-5} + \mathcal{O}(R^{-6}),
\end{equation}
which leads to explicit expressions for the dispersion coefficients. In this work we will truncate the expansion at the first term and focus on the $C_6$ dispersion coefficients. As derived in the supplementary material of ref \citenum{KooGor-JPCL-19}, the variational equation for $C_6$ is given in terms of matrices $\tau_{ij}$, $S_{ij}$ and $P_{ij}$,
\begin{subequations}
\begin{align}
\tau_{ij} &= \int d\rv \rho(\rv) \nabla b_i(\rv) \cdot \nabla b_j(\rv)\\
S_{ij} &= \int d\rv \rho(\rv) b_i(\rv) b_j(\rv)\\
P_{ij} &= \int d\rv_{1A} \int d\rv_{2A} P_2(\rv_{1A}, \rv_{2A}) b_i(\rv_{1A}) b_j(\rv_{2A}),
\end{align}
\end{subequations}
defined for both subsystems $A$ and $B$. The terms $\tau_{ij}$ and $S_{ij}$ depend only on the one-electron density, but $P_{ij}$ depends on the pair density of each monomer. The monomer--monomer interaction is described by the $w_{ij}$ matrix,
\begin{equation}
w_{ij} = (\textbf{d}_{i}^A + \textbf{D}_{i}^A)(\textbf{d}_{i}^B + \textbf{D}_{i}^B).
\end{equation}
The dipole--dipole interaction terms $d_{i}, D_{i}$ consist of electron density and pair density mediated parts,
\begin{subequations}
\begin{align}
d_{i} &= \int d\rv \rho(\rv) b_i(\rv) \rv_{1A}\\
D_{i} &= \int d\rv_{1A} \int d\rv_{2A} P_2(\rv_{1A}, \rv_{2A}) b_i(\rv_{2A}) \rv_{1A}.
\end{align}
\end{subequations}
The matrix $\tau_{ij}^{A/B}$ is diagonalized with $S_{ij}^{A/B}+P_{ij}^{A/B}$ as a metric through a generalized eigenvalue problem, with $w_{ij}$ also transformed accordingly. This leads to a simple form for the dispersion coefficient $C_6$:
\begin{equation}
\label{eqn:C6coeff}
C_6^\text{FDM}[\rho^A, \rho^B, P_2^A, P_2^B] = 2\sum_{ij}\frac{w_{ij}^2}{\tau^A_i+\tau_j^B}.
\end{equation}


\subsection{Exchange-correction to the Hartree--Fock dispersion coefficients}

While the diagonal of the many-body density matrix is fixed in the FDM procedure, the off-diagonal elements may change when the systems $A$ and $B$ interact. If the Hartree--Fock pair density is used instead of the exact pair density for each monomer, a change in these off-diagonal terms affects the exchange energy and this change should be included. We can derive a correction to the FDM procedure described above by looking at how the constraint changes the 1-RDM of the monomer $A$ (the same treatment applied for monomer $B$ as well) and how this in turn causes a change in the exchange energy via the exchange kernel.

In the FDM framework, the 1-RDM of monomer $A$ can be written as
\begin{align*}
\gamma^A_R(\xv'_A, \xv_A) = N_A\int d\xv_{A_2}\ldots d\xv_{A_N} &\Psi^A*_0(\xv'_A, \xv_{A_2},\ldots)\Psi^A*(\xv_A, \xv_{A_2},\ldots) \\&\times\int d\underline{\xv}_B|\Psi^B_0(\underline{\xv}_B|^2  \sqrt{(1+\underline{J'})(1+\underline{J})}
\end{align*}
where we have denoted
\begin{subequations}
\begin{align}
\underline{J'} &= \sum_{j\in B} J_R(\rv', \rv_j) + \sum_{i\neq 1 \in A, j\in B} J_R(\rv_i, \rv_j)  \\
\underline{J} &= \sum_{j\in B} J_R(\rv, \rv_j) + \sum_{i\neq 1 \in A, j\in B} J_R(\rv_i, \rv_j).
\end{align}
\end{subequations}
If we expand the square root up to the second order in $\underline{J}, \underline{J'}$, we obtain
\begin{equation}
\sqrt{(1+\underline{J'})(1+\underline{J})} = 1 + \frac{1}{2}(\underline{J} + \underline{J'}) - \frac{1}{8} (\underline{J} - \underline{J'})^2 + \ldots
\end{equation}
The zeroth order terms corresponds to the $\gamma_0^A(\xv_A', \xv_A)$, i.e. the Hartree--Fock 1-RDM of the isolated monomer. The linear term vanishes due to the constraints. The first nontrivial term in the expansion is the second order quadratic term. Using the definitions for $\underline{J}$ and $\underline{J'}$ above, we obtain
\begin{equation}
\underline{J} - \underline{J'} = \sum_{j\in B} \left(J_R(\rv, \rv_j) - J_R(\rv', \rv_j)\right).
\end{equation}
If we insert the one-body function expansion of $J$ (equation \ref{eqn:J-expansion}), we get as the FDM 1-RDM
\begin{equation}
\gamma^A_R(\xv'_A, \xv_A) = \gamma^A_0(\xv'_A, \xv_A)\left(1+Q_R^A(\rv, \rv')\right)
\end{equation}
where
\begin{equation}
Q_R^A(\rv, \rv') = -\frac{1}{8} \sum_{ijkl} c_{ij}c_{kl}\left(b_i^A(\rv)b_k^A(\rv) + b_i^A(\rv')b_k^A(\rv') - b_i^A(\rv)b_k^A(\rv')-b_i^A(\rv')b_k^A(\rv)\right) (S_{ij}^B + P_{ij}^B) 
\end{equation}
The change in the nondiagonal terms leads to a change in the expectation value of the exchange operator $\hat{K}^A$, which has a kernel
\begin{equation}
\hat{K}^A = -\frac{\gamma_0^A(\xv,\xv')}{|\rv-\rv'|}.
\end{equation}
We then obtain as a change in the expectation value,
\begin{equation}
\Delta K^A = \frac{1}{8} \sum_{ijkl} c_{ij}c_{kl}K^A_{ik}(S^B_{jl} + P^B_{jl})
\end{equation}
where 
\begin{equation}
\begin{aligned}
K_{ik}^A &= \int d\xv \int d\xv' \frac{|\gamma_0^A(\xv,\xv')|^2}{|\rv-\rv'|} \left(b_i^A(\rv)b_k^A(\rv) + b_i^A(\rv')b_k^A(\rv') - b_i^A(\rv)b_k^A(\rv')-b_i^A(\rv')b_k^A(\rv)\right).
\end{aligned}
\end{equation}
This exchange contribution is added to the kinetic energy contribution, $\tau_{ik} \rightarrow \tau_{ik} + K_{ik}$. Since the correction is quadratic, the increase in magnitude of the denominator in eqn \ref{eqn:C6coeff} will help to offset the overestimation of the dispersion coefficients observed for Hartree--Fock dispersion coefficients\cite{KooWecGor-JCTC-21}.

\subsection{Becke--Roussel exchange-hole}

The FDM method reproduces accurate dispersion coefficients\cite{KooWecGor-JCTC-21} when a correlated pair density (MP2 or CCSD) is used to compute the pair density mediated matrix elements. However, obtaining an accurate pair density is computationally expensive and, for developing practical applications, it is interesting to see how an approximation to pair density built on the ground state electron density would perform in the FDM procedure. Therefore, we will test the performance of the real space exchange hole model proposed by Becke--Roussel\cite{BecRou-PRA-89} (BR) in computing the pair density mediated matrix elements.

Consider, that the pair density of a system is related to the exchange--correlation hole $h_{xc}$ by
\begin{equation}
P_2(\rv_{1A}, \rv_{2A}) = \rho(\rv_{1A})\rho(\rv_{2A}) +\rho(\rv_{1A}) h_{xc} (\rv_{1A}, \rv_{2A}).
\end{equation}
We can then rewrite the pair density mediated terms $P_{ij}$ (overlap) and $D_{i}$ (dipole moment) using the xc-hole,
\begin{subequations}
\begin{align}
P_{ij} &= \int d\rv_{1A} \rho(\rv_{1A}) b_i(\rv_{1A})\int d\rv_{2A} h_{xc} (\rv_{1A}, \rv_{2A}) b_j(\rv_{2A})\\
D_{i,e} &= \int d\rv_{1A} \rho(\rv_{1A})  \rv_{1A} \int d\rv_{2A} h_{xc}(\rv_{1A}, \rv_{2A}) b_j(\rv_{2A}).
\end{align}
\end{subequations}
The exchange--correlation hole is often decomposed into a sum of exchange and correlation,
\begin{equation}
h_{xc} (\rv_{1A}, \rv_{2A}) = h_{x} (\rv_{1A}, \rv_{2A}) + h_{c} (\rv_{1A}, \rv_{2A}).
\end{equation}
For our purposes here, we will approximate the exchange-hole using the Becke--Roussel model for the exchange-hole and neglect the correlation term. 

The Becke--Roussel exchange-hole is based on generalizing the shape of the exchange hole of an hydrogenic atom to other systems. An exponential function is placed at a distance $b(\rv)$ from the position $\rv$ of the $\sigma$ reference electron and properly normalized,\cite{BecRou-PRA-89}
\begin{equation}
h_{x,\sigma}^\text{BR}(\rv_{1}, \rv_{2}) = - \frac{a^3(\rv_1)}{8\pi} e^{-a(\rv_1) |\rv_2 - \rv_1 + \textbf{b}(\rv_1)|}.
\end{equation}
The BR exchange hole depends on the parameters $a(\rv), {b}(\rv)$ that are determined by requiring that the exchange-hole satisfies the Taylor expansion of the exact spherically averaged exchange hole near the reference point up to the second order:
\begin{align*}
h_{X\sigma} (\rv, s) &= -\rho_\sigma + \frac{s^2}{6} \left(2 \tau_\sigma - \frac{(\nabla\rho_\sigma)^2}{2\rho_\sigma} - \nabla^2 \rho_\sigma\right) + \ldots
\end{align*}
where $\tau_\sigma = \sum_i|\nabla\psi_{i,\sigma}|^2$, $\rho_\sigma$ is the spin-density and $\psi_{i,\sigma}$ are the Kohn--Sham spin-orbitals. Requiring that the BR-hole is equal to the exact exchange-hole in the second-order, one obtains two equations for $a(r), {b}(r)$,
\begin{align}
a^3e^{-ab} &= \rho_\sigma 8\pi \label{eqn:br_params1}\\
a^2b-2a &= \frac{6bQ_\sigma}{\rho_\sigma}. \label{eqn:br_params2}
\end{align}
Proynov \emph{et al.}\cite{ProGanKon-CPL-08} have derived an analytical representation for equations \ref{eqn:br_params1} and \ref{eqn:br_params2}, allowing for an efficient evaluation of the parameters $a(r)$ and $b(r)$ without the need to numerically solve them for each $r$.

In their original paper\citep{BecRou-PRA-89}, Becke and Roussel looked for an approximation for the spherically averaged exchange hole to computed the exchange energy, since the Coulomb interaction depends only on the electron--electron distance $s = |\rv_1-\rv_2|$. 
However, in this work we consider a more general form of the BR exchange hole. If we do not take a spherical average over $s$, we are left with a vector quantity $\textbf{b}(\rv)$ that we need to choose a direction for\cite{GorAngSav-CJC-09}. For a spherically symmetric atom, the only choice is to have $\textbf{b}(\rv)$ point towards the nucleus, parallel to $\rv$. Then we may rewrite the BR hole as
\begin{equation}
h_{x,\sigma}^\text{BR}(\rv_{1}, \rv_{2}) = - \frac{a^3(\rv_1)}{8\pi} e^{-a(\rv_1) \left|\left(\frac{b(r_1)}{r_1} -1 \right) \rv_1 + \rv_2 \right|}.
\end{equation}
This expression for the BR-hole was used by Gori--Giorgi \emph{et al.} in their study\cite{GorAngSav-CJC-09} on the charge reconstruction using approximate exchange--correlation-holes.


\section{Computational Details}

\begin{figure}
\begin{center}
\includegraphics[width=0.4\textwidth]{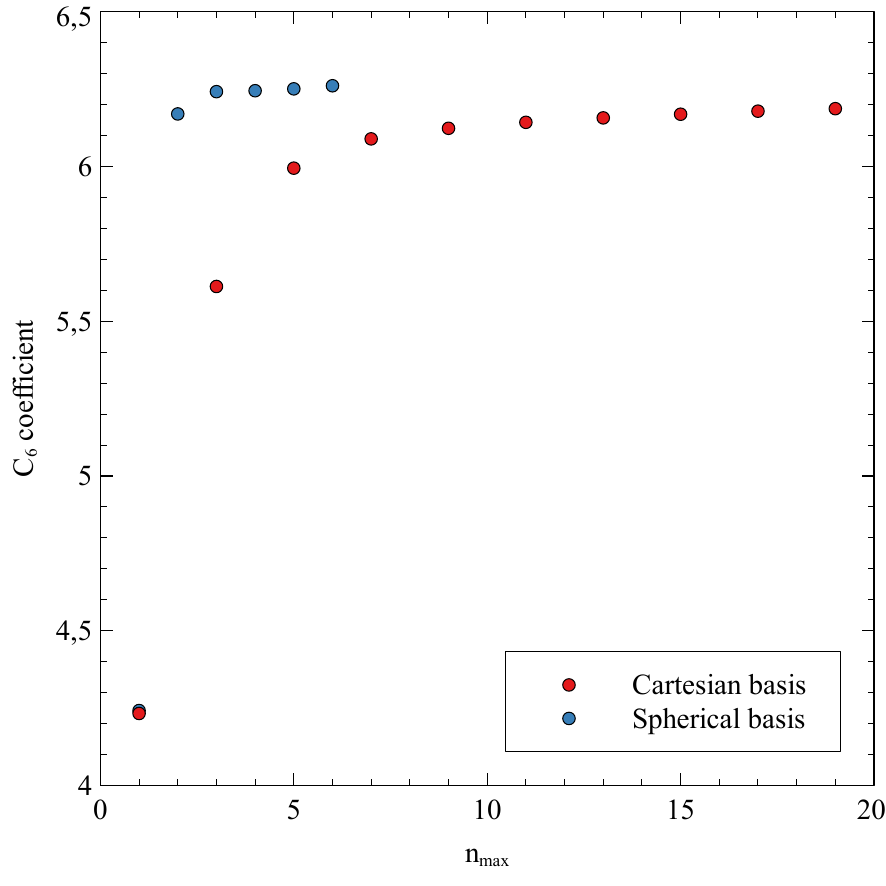}
\end{center}
\caption{The convergence of the dispersion coefficient for Neon as a function of highest order of $r$ in the $b$ basis using the CCSD pair density. The Cartesian monomial basis does not include the odd powers of $r$. The spherical $b$-functions converge rapidly, but the monomial basis requires very high orders of $n$ and still remains below the range achieved by the spherical basis.}
\label{fig:convergence}
\end{figure}

In our previous work\cite{KooGor-JPCL-19, KooWecGor-JCTC-21}, a Cartesian monomial basis was used for the $b$-functions that allowed for a fast computation with the implementation used. Here, a set of spherical $b$-functions will be used, which is more applicable for the atomic systems:
\begin{equation}
b_i(\rv) = r^i Y_{1m}(\theta, \phi),
\end{equation}
where $i$ is a positive integer. We set $m=0$ as $m=\pm 1$ are redundant. In the spherical basis, the dispersion coefficients converge significantly faster when compared to a monomial basis (see figure \ref{fig:convergence}). The integration of the multipole moments was done numerically on a grid with $X\times Y$ grid points, where $X$ denotes the grid points on a radial grid and $Y$ the grid points of a Lebedev--Laikov grid. 

\subsection{Exchange-correction to the Hartree--Fock dispersion coefficients}

For the evaluation of the exchange correction, atomic densities and orbitals were obtained on a grid from PySCF. The correction 
\begin{equation}
\begin{aligned}
K_{ik}^A &= \sum_{pqrs} \gamma_{pq}\gamma_{rs} \left<pq| rs\right>_{ik}\\
\left<pq| rs\right>_{ik} &= \int d\xv \int d\xv' \frac{\phi_p^*(\xv)\phi_r(\xv)\phi_q^*(\xv')\phi_s(\xv')}{|\rv-\rv'|} \left(b_i^A(\rv)b_k^A(\rv) + b_i^A(\rv')b_k^A(\rv') - b_i^A(\rv)b_k^A(\rv')-b_i^A(\rv')b_k^A(\rv)\right)
\end{aligned}
\end{equation}
was computed numerically and the exchange correction $K_{ik}$ was added to the kinetic energy matrix $\tau_{ik}$. A numerical grid of $20\times26$ grid for both $\xv_1$ and $\xv_2$ was used. A sparse Lebedev--Laikov grid is used to evaluate the correction terms as the exchange correction appears to be insensitive to the size of the angular part of the grid (see supplementary information for details on the convergence of the calculations).

\subsection{Becke--Roussel multipole moments}

The atomic densities and derivatives of orbitals and density used for the Becke--Roussel integrals were based on a CCSD ground state density of a given system. All the quantities were obtained on a grid from PySCF. An analytical expression by Proynov \emph{et al.}\cite{ProGanKon-CPL-08} was used to calculate the parameters of the BR hole. 
For molecules, the electron density was spherically averaged, with origin placed on the center of mass of the molecule. 
For computing the BR multipole integrals, a grid of $35\times350$ grid for both $\xv_1$ and $\xv_2$ was deemed necessary (see supplementary information for further details).

Using an approximate exchange--correlation hole has the disadvantage that the multipole moments are not in general symmetric, i.e. $P_{ij} \neq P_{ji}$. This asymmetry is due to the fact that the exact exchange--correlation hole satisfies the identity, $\int d\rv \rho(\rv)h_{xc}(\rv, \rv') = -\rho(\rv')$ while the approximate exchange--correlation holes in general do not\cite{GorAngSav-CJC-09}. To correct for this, we take an average, $\tilde{P}_{ij} = \frac{P_{ij}+P_{ji}}{2}$, of the integrals in our computations.

\section{Results}\label{sec:results}

\begin{figure}
\begin{center}
\includegraphics[width=0.4\textwidth]{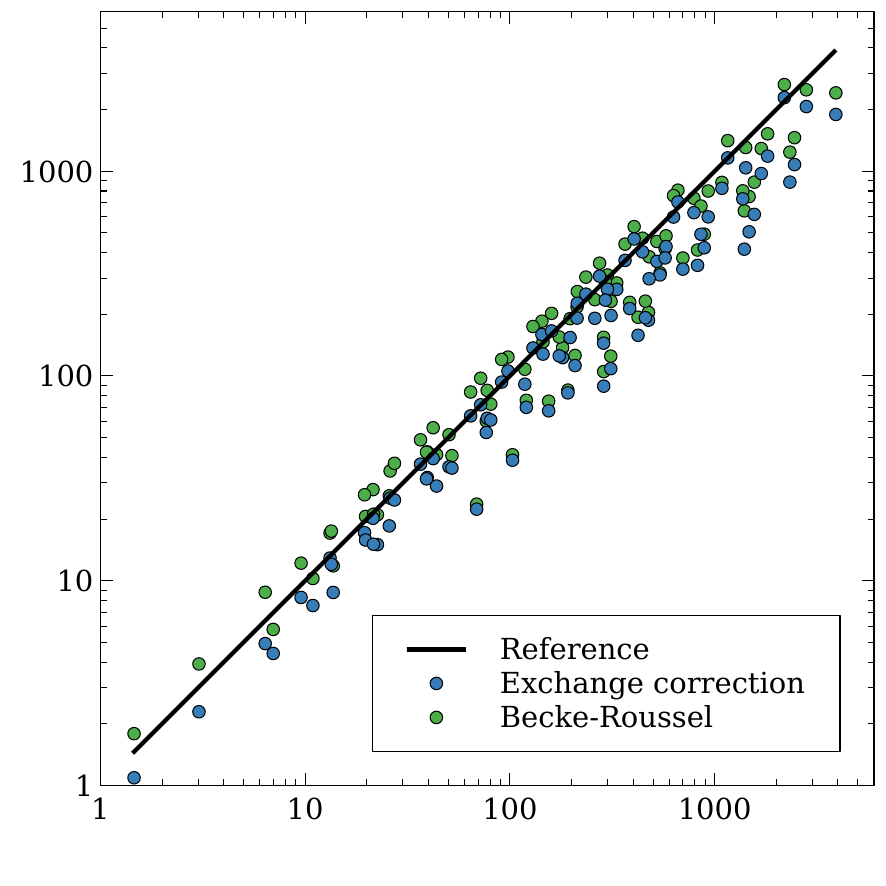}
\includegraphics[width=0.4\textwidth]{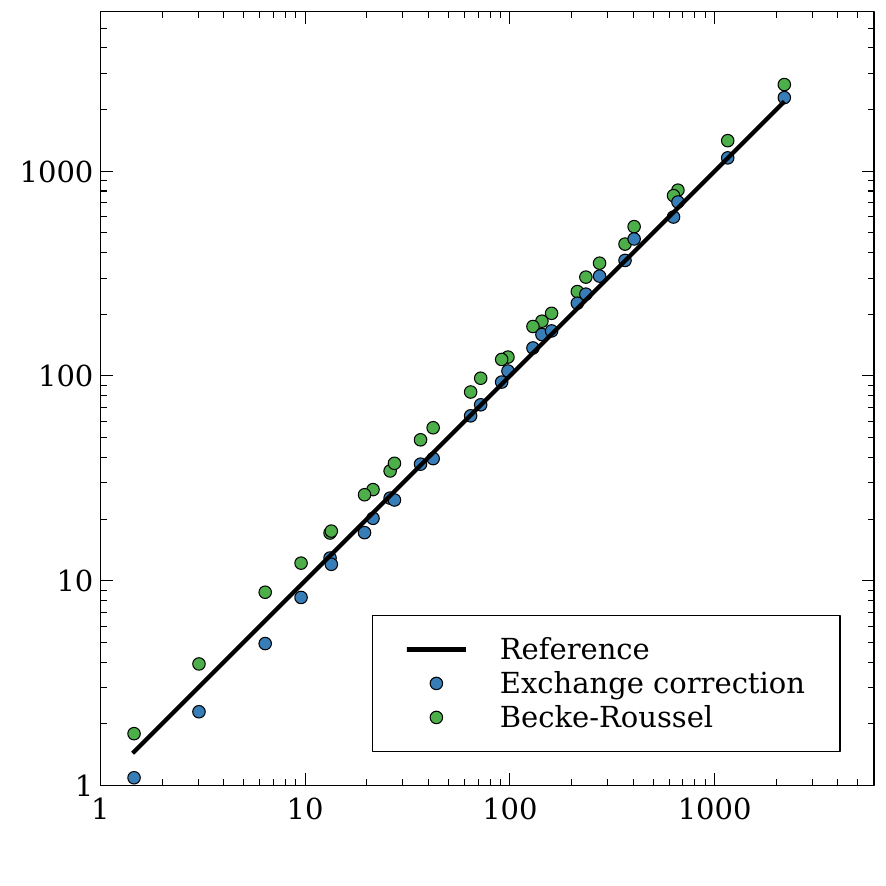}
\end{center}
\caption{Left: Dispersion coefficients $C_6$ for 91 pairs of closed and open shell atoms and ions computed using either the Becke--Roussel exchange-hole approximation or the Hartree--Fock pair density with the exchange-correction. Right: Subset of the computed dispersion coefficients consisting only of the 28 closed-shell species. Reference data obtained from ref \citenum{JiaMitCheBro-ATDNDT-15}.}
\label{fig:c6results}
\end{figure}

The performance of the exchange-correction and the BR exchange-hole are compared with closed- and open-shell atoms and ions for which there is accurate reference data available\cite{JiaMitCheBro-ATDNDT-15}. The closed-shell atoms consist of noble gas atoms from He to Kr and alkaline earth metals from Be to Ca and the open shell atoms and ions consist of alkali metals and alkaline earth metal ions. The dispersion coefficients for obtained from the two approaches are depicted in figure \ref{fig:c6results}.

\subsection{Exchange-correction to the Hartree--Fock dispersion coefficients}

The dispersion coefficient computed using the Hartree--Fock method with and without the exchange correction and using the CCSD pair density for the closed and open shell atoms and ions are presented Table \ref{tbl:HF_all_atoms} open shell atoms and ions. In addition, dispersion coefficients for different closed shell mixed pairs computed with the CCSD pair density, the Hartree--Fock pair density with and without the exchange correction, and Kohn--Sham pair density with the exchange correction  using both LDA and PBE functionals, are presented in Table \ref{tbl:all_closed_shell_atoms}. 

\begin{table}
\caption{Dispersion coefficients for all closed-shell atom pairs computed using Becke--Roussel exchange hole without scaling ($d=1.00$) and with the scaling by 25\% ($d=1.25$). All values are computed using a def2-tzvpp basis and a dispersal basis of $n=10$. }
\label{tbl:HF_all_atoms}
\begin{tabular}{l r r r r}
Pair & Ref	&	Hartree--Fock &	HF K-corrected & CCSD\\
\hline
Closed shell atoms\\
He	&	1.46	&	1.62	&	1.09	&	1.43	\\
Be	&	213.41	&	468.20	&	226.46	&	168.79	\\
Ne	&	6.38	&	6.88	&	4.93	&	6.26	\\
Mg	&	629.59	&	1241.39	&	596.84	&	518.14	\\
Ar	&	64.30	&	96.87	&	63.82	&	58.84	\\
Ca	&	2188.20	&	5004.31	&	2285.81	&	1799.56	\\
Kr	&	129.56	&	210.14	&	136.92	&	122.19	\\
\textbf{MAPE:}	&	&	\textbf{68.1}\%	&	\textbf{10.1}\%	 & \textbf{10.7}\%\\
\textbf{AMAX:}	&	&	\textbf{128.7}\%	&	\textbf{25.4}\%&	\textbf{20.9}\%\\
\hline
Open shell atoms and ions\\
\ce{Li}	&	1395.80	&	1028.51	&	415.80	&	985.32	\\
\ce{Be+}	&	68.80	&	40.77	&	22.34	&	39.81	\\
\ce{Na}	&	1561.60	&	1449.82	&	615.05	&	1206.25	\\
\ce{Mg+}	&	154.59	&	120.25	&	67.52	&	109.32	\\
\ce{K}	&	3906.30	&	4464.31	&	1891.80	&	2959.24	\\
\ce{Ca+}	&	541.03	&	562.95	&	311.22	&	380.81	\\
\textbf{MAPE:}	&	&	\textbf{19.1}\%	&	\textbf{58.1}\%&	\textbf{29.6}\%	\\
\textbf{AMAX:}	&	&	\textbf{40.7}\%	&	\textbf{70.2}\%&	\textbf{42.1}\%
\end{tabular}
\end{table}

In general, the Hartree--Fock pair density results in an overestimation of the dispersion coefficients\cite{KooWecGor-JCTC-21} with a mean absolute percentage error (MAPE) of 68.1\%. However, when the exchange correction is added to the kinetic energy, the magnitude of the denominator in eqn \ref{eqn:C6coeff} increases, decreasing the overall value of the dispersion coefficients. The resulting exchange corrected dispersion coefficients line very well with the reference data and the MAPE decreases significantly to 10.1\%. This is close to the error obtained using the CCSD pair density with a MAPE of 10.6\% for the closed-shell atoms. 

The FDM procedure tends to perform poorly on open-shell systems, resulting in a dispersion coefficient well below the reference values. Due to the intrinsic overestimation of the dispersion coefficients, the Hartree--Fock pair density results in relatively good performance with an MAPE of 19.1\%. This should be compared to the 29.6\% error obtained when using the correlated CCSD pair density. When the exchange--correction is applied to open shell systems, the results are much worse than without the correction, leading to an MAPE of 58.2\%. 

In addition, we evaluated the performance of the Kohn--Sham pair density with and without the exchange-correction (see supplementary data). The dispersion coefficients were computed using the LDA\cite{lda} and PBE\cite{pbe} functionals. Without the exchange-correction, LDA performs worse (MAPE 78.8\%) than Hartree--Fock, while PBE performs slightly better (MAPE 56.1\%). With the exchange-correction included, the MAPE of both LDA and PBE are reduced significantly to 11.1\% and 11.4\%, respectively, close to what is obtained for Hartree--Fock with exchange-correction.

\begin{table}
\caption{Dispersion coefficients for all closed-shell atom pairs computed using Hartree--Fock, CCSD and Kohn--Sham pair densities using LDA and PBE functionals. All values are computed using a def2-tzvpp basis and a dispersal basis of $n=6$. For Hartree--Fock and Kohn--Sham pair densities, the exchange-correction is denoted by "-K". }
\label{tbl:all_closed_shell_atoms}
\begin{tabular}{l r r r r r r r}
Pair & Ref	&	HF (n=6)	&	HF-K	&	CCSD &	LDA-K	 &	PBE-K	\\
\hline
He -- He & 1.46 & 1.62 & 1.09 & 1.43 & 1.36 & 1.19 \\
He -- Be & 13.23 & 20.43 & 12.90 & 12.60 & 15.25 & 13.13 \\
He -- Ne & 3.03 & 3.30 & 2.29 & 2.96 & 2.84 & 2.59 \\
He -- Mg & 21.45 & 31.59 & 20.16 & 20.83 & 22.35 & 19.88 \\
He -- Ar & 9.55 & 12.40 & 8.28 & 9.10 & 9.71 & 8.74 \\
He -- Ca & 36.58 & 57.77 & 37.07 & 36.06 & 40.06 & 35.78 \\
He -- Kr & 13.42 & 18.04 & 12.02 & 13.01 & 14.28 & 12.82 \\
Be -- Be & 213.41 & 468.20 & 226.46 & 168.79 & 249.82 & 207.32 \\
Be -- Ne & 26.00 & 37.94 & 25.30 & 24.48 & 30.01 & 26.77 \\
Be -- Mg & 364.89 & 758.95 & 366.45 & 293.83 & 373.57 & 321.99 \\
Be -- Ar & 97.82 & 174.02 & 105.65 & 86.65 & 116.65 & 102.69 \\
Be -- Ca & 661.61 & 1487.44 & 706.62 & 535.98 & 698.37 & 603.61 \\
Be -- Kr & 143.32 & 267.43 & 159.24 & 128.54 & 178.94 & 156.57 \\
Ne -- Ne & 6.38 & 6.88 & 4.93 & 6.26 & 6.06 & 5.69 \\
Ne -- Mg & 42.18 & 58.60 & 39.46 & 40.39 & 43.98 & 40.51 \\
Ne -- Ar & 19.50 & 24.74 & 17.16 & 18.66 & 20.08 & 18.69 \\
Ne -- Ca & 71.99 & 106.78 & 72.26 & 69.67 & 78.57 & 72.69 \\
Ne -- Kr & 27.30 & 35.72 & 24.77 & 26.54 & 29.35 & 27.27 \\
Mg -- Mg & 629.59 & 1241.39 & 596.84 & 518.14 & 560.26 & 502.22 \\
Mg -- Ar & 159.79 & 270.28 & 165.87 & 144.32 & 171.31 & 156.00 \\
Mg -- Ca & 1158.00 & 2466.04 & 1160.93 & 958.20 & 1052.78 & 947.32 \\
Mg -- Kr & 234.94 & 416.86 & 250.78 & 215.08 & 263.34 & 238.47 \\
Ar -- Ar & 64.30 & 96.87 & 63.82 & 58.84 & 70.22 & 64.79 \\
Ar -- Ca & 274.03 & 498.14 & 307.12 & 251.69 & 308.84 & 282.41 \\
Ar -- Kr & 91.13 & 142.32 & 93.31 & 84.65 & 104.06 & 95.67 \\
Ca -- Ca & 2188.20 & 5004.31 & 2285.81 & 1799.56 & 1997.46 & 1804.02 \\
Ca -- Kr & 404.19 & 772.25 & 466.30 & 376.78 & 476.83 & 433.50 \\
Kr -- Kr & 129.56 & 210.14 & 136.92 & 122.19 & 154.91 & 141.87 \\
\textbf{MAPE:}	&	&	\textbf{63.8}\%	&	\textbf{8.1}\%	&	\textbf{8.3}\% & \textbf{10.2}\%  &  \textbf{7.2}\% \\
\textbf{AMAX:}	&	&	\textbf{128.7}\%	&	\textbf{25.4}\%	&	\textbf{20.9}\% & \textbf{24.9}\%  & \textbf{20.2}\% 
\end{tabular}
\end{table}

\subsection{Becke--Roussel exchange-hole}

While the exchange-correction is relatively stable numerically and converges for a small angular grid, the Becke--Roussel exchange-hole requires a dense angular grid. Also, due to the approximate nature of the exchange-hole, the overlap matrix produced by the BR multipole moment integrals is not guaranteed to be positive definite, and for $n$ above ten, the moment integrals could not be computed for all elements due to the overlap matrix being singular.

The dispersion coefficients for closed shell atoms computed using the Becke--Roussel exchange-hole are tabulated in Table \ref{tbl:BR_all_atoms} and \ref{tbl:BR_open_shell}. The Becke--Roussel model overestimates the dispersion coefficients of the closed-shell atoms and significantly underestimates the dispersion coefficients of the open-shell systems. The MAPE for closed-shell atoms is 26.8\% and 45.4\% for open-shelled ions and atoms. 

The multipole moment integrals obtained from the BR exchange-hole should be compared to those obtained using an accurate pair density. The ratios of the multipole moment integrals calculated using the BR exchange hole and the CCSD pair density for selected atoms are shown in table S4. For small values of $n$, the BR integrals are in general smaller than those obtained using the CCSD pair density. Since the low $n$ terms have the largest weight in the overall dispersion coefficient, the magnitude of the dispersion coefficients tends to be overestimated. At high values of $n$, the trend reverses and the magnitude of the multipole moment obtained from BR exchange-hole are significantly larger than those obtained from CCSD. 

\begin{table}
\caption{Dispersion coefficients for all closed-shell atom pairs computed using Becke--Roussel exchange hole without scaling ($d=1.00$) and with the scaling by 25\% ($d=1.25$). All values are computed using a def2-tzvpp basis and a dispersal basis of $n=10$. }
\label{tbl:BR_all_atoms}
\begin{tabular}{l r r r}
Pair & Ref	&	BR ($d=1.00$) &	BR ($d=1.25$)	\\
\hline
Closed shell atoms\\
He	&	1.46	&	1.79	&	1.83	\\
Be	&	213.41	&	258.10	&	228.93	\\
Ne	&	6.38	&	8.78	&	6.47	\\
Mg	&	629.59	&	759.50	&	670.28	\\
Ar	&	64.30	&	83.44	&	51.34	\\
Ca	&	2188.20	&	2647.71	&	2235.58	\\
Kr	&	129.56	&	174.28	&	103.86	\\
\textbf{MAPE:}	&	&	\textbf{26.7}\%	&	\textbf{8.5}\%	\\
\textbf{AMAX:}	&	&	\textbf{37.6}\%	&	\textbf{20.2}\%\\
\hline
Open shell atoms\\
Li	&	1395.80	&	639.73	&	561.86	\\
Be+	&	68.80	&	23.64	&	20.87	\\
Na	&	1561.60	&	883.62	&	783.09	\\
Mg+	&	154.59	&	75.27	&	63.40	\\
K	&	3906.30	&	2411.49	&	2045.43	\\
Ca+	&	541.03	&	318.43	&	246.50	\\
\textbf{MAPE:}	&	&	\textbf{49.0}\%	&	\textbf{56.7}\%	\\
\textbf{AMAX:}	&	&	\textbf{65.6}\%	&	\textbf{69.7}\%
\end{tabular}
\end{table}

\begin{table}
\caption{Dispersion coefficients for all closed-shell atom pairs computed using Becke--Roussel exchange hole without scaling ($d=1.00$) and with the scaling by 25\% ($d=1.25$). All values are computed using a def2-tzvpp basis and a dispersal basis of $n=10$. }
\label{tbl:BR_all_atoms}
\begin{tabular}{l r r r}
Pair & Ref	& BR ($d=1.00$) &	BR ($d=1.25$)	\\
\hline
He	--	He	&	1.46	&	1.79	&	1.83	\\
He	--	Be	&	13.23	&	17.09	&	16.46	\\
He	--	Ne	&	3.03	&	3.92	&	3.38	\\
He	--	Mg	&	21.45	&	27.84	&	26.68	\\
He	--	Ar	&	9.55	&	12.18	&	9.67	\\
He	--	Ca	&	36.58	&	48.74	&	45.50	\\
He	--	Kr	&	13.42	&	17.46	&	13.68	\\
Be	--	Be	&	213.41	&	258.10	&	228.93	\\
Be	--	Ne	&	26.00	&	34.33	&	27.76	\\
Be	--	Mg	&	364.89	&	440.38	&	389.53	\\
Be	--	Ar	&	97.82	&	123.57	&	90.04	\\
Be	--	Ca	&	661.61	&	807.42	&	697.96	\\
Be	--	Kr	&	143.32	&	184.99	&	133.39	\\
Ne	--	Ne	&	6.38	&	8.78	&	6.47	\\
Ne	--	Mg	&	42.18	&	55.82	&	44.87	\\
Ne	--	Ar	&	19.50	&	26.28	&	17.81	\\
Ne	--	Ca	&	71.99	&	97.43	&	76.19	\\
Ne	--	Kr	&	27.30	&	37.43	&	24.94	\\
Mg	--	Mg	&	629.59	&	759.50	&	670.28	\\
Mg	--	Ar	&	159.79	&	202.14	&	146.45	\\
Mg	--	Ca	&	1158	&	1408.76	&	1215.73	\\
Mg	--	Kr	&	234.94	&	303.76	&	217.86	\\
Ar	--	Ar	&	64.3	&	83.44	&	51.34	\\
Ar	--	Ca	&	274.03	&	355.29	&	250.58	\\
Ar	--	Kr	&	91.13	&	120.35	&	72.84	\\
Ca	--	Ca	&	2188.2	&	2647.71	&	2235.58	\\
Ca	--	Kr	&	404.19	&	535.82	&	374.33	\\
Kr	--	Kr	&	129.56	&	174.28	&	103.86	\\
\textbf{MAPE:}	&	&	\textbf{28.8}\%	&	\textbf{10.4}\%	\\
\textbf{AMAX:}	&	&	\textbf{37.6}\%	&	\textbf{25.0}\%
\end{tabular}
\end{table}

\subsection{Scaling the Becke--Roussel terms}

Since the pair density mediated terms from the BR exchange hole are underestimated compared to terms computed with the more accurate CCSD pair density, assuming the underestimation is consistent enough, one might consider improving the performance of the BR exchange-hole simply by scaling the pair density mediated matrix elements by a scaling factor $d$. Assuming the pair density mediated matrix elements are positive, scaling them decreases the computed dispersion coefficients. The values for the dispersion coefficient with different scaling parameter values are presented in table S9. The optimal scaling parameter for the set of closed-shell atoms is $d=1.15$, where MAPE is decreased to 10.7\%. With a scaling parameter of $d=1.25$, the MAPE is slightly higher, 11.8\%. For open-shell atoms and ions, scaling the moments upward does not improve the performance since the dispersion coefficients are underestimated by the FDM procedure. 

\begin{figure}
\begin{center}
\includegraphics[width=0.4\textwidth]{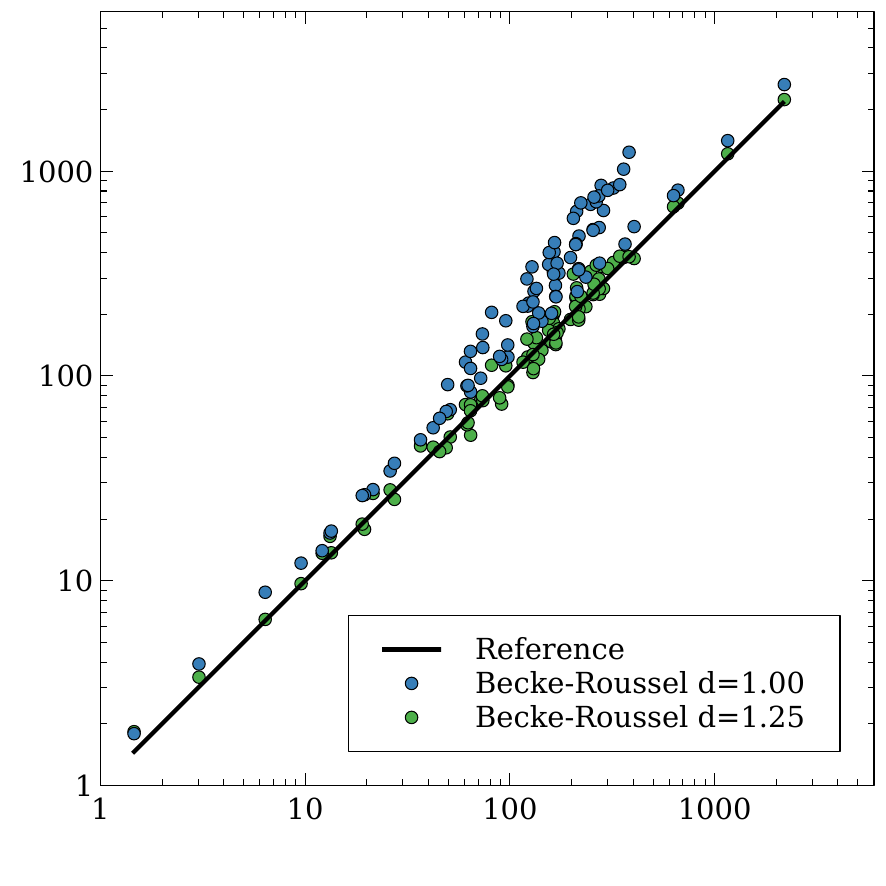}
\end{center}
\caption{Dispersion coefficients $C_6$ for all the closed shell atoms and molecules computed using the Becke--Roussel exchange-hole with no scaling ($d=1.00$) or with 25\% scaling ($d=1.25$). The data is also available in Tables \ref{tbl:BR_all_atoms}, S10 and S11.}
\label{fig:br_scaling}
\end{figure}

In addition to the set of closed-shell atoms we computed the dispersion coefficient for a set of small molecules using the BR exchange-hole. These results are presented in table S10. For molecules, the overestimation of the dispersion coefficients is more pronounced than for the atoms. A scaling parameter of $d=1.25$ decreases the MAPE from 104.9\% down to 14.6\%. Accounting for different possible mixed combinations with reference data available (see Table S11 in the Supplementary Information), the MAPE decreases from 107.1\% down to 12.7\%. The computed dispersion coefficients with respect to the reference data for all the closed shell systems with and without the scaling are presented in Figure \ref{fig:br_scaling}. 

\section{Conclusions and Perspectives}

In this work, an exchange-correction to the FDM procedure was presented. This correction improves the accuracy of the FDM method when a single reference pair density is used. The performance of the Becke--Roussel exchange-hole was also studied for approximating the pair density mediated integrals. 

We find that the exchange-correction significantly improves the performance of the Hartree--Fock pair density, bringing the mean absolute percentage error for closed-shell atoms down from 62\% to 9\%, close to that obtained using the correlated CCSD pair density (8\%). These results show the FDM methods yields dispersion coefficients for closed-shell atoms on par with highly correlated methods when using only a single Slater determinant. The exchange-correction also allows the accurate estimation of dispersion coefficients using density functional theory from the Kohn--Sham pair density. The method is not highly sensitive to the level of the approximate exchange--correlation functional used. 

The approximate real space model for the exchange-hole proposed by Becke and Roussel was used to compute the pair density mediated terms. While the method improved the results for closed-shell atoms compared to dispersion coefficient obtained using the Hartree--Fock pair density without the exchange--correction, the mean absolute percentage error was still high (26.7\%). When the pair density mediated overlap and dipole moment integrals from the Becke--Roussel model were compared to those computed using the accurate CCSD pair density, the Becke--Roussel model integrals for small $n$ were smaller in magnitude than the CCSD integrals. A simple scaling of the Becke--Roussel integrals significantly improved the performance of the approximate exchange hole model, that worked for both the closed shell atoms but also closed shell molecules.

For open-shell ions and atoms the exchange--correction performed worse than the CCSD reference (MAPE 58.0\% vs 29.4\%). The dispersion coefficients for the open-shell systems are large in magnitude and are underestimated even when the Hartree--Fock pair density is used without the exchange-correction (MAPE 19.3\%). Also the Becke--Roussel exchange-hole underestimates the dispersion coefficients (MAPE of 49.0\%). Thus, while the FDM procedure can obtain accurate dispersion coefficients for the closed-shell atoms and molecules, the open-shell systems remains a weak point for the FDM method.

Overall, our results show that, when employing the exchange correction, the FDM procedure is able to produce accurate dispersion coefficients for closed-shell atoms using a single reference wave function alone. The error for the obtained dispersion coefficients is close to that obtained for the accurate CCSD pair density. The pair density mediated terms approximated using the Becke--Roussel exchange hole results in comparable performance, when the pair density mediated terms are scaled up by about 25\%. 

\section*{Acknowledgements}

T.W. acknowledges Derk Kooi and Paola Gori-Giorgi from the Vrije Universiteit Amsterdam for collaboration and helpful discussions. T.W. acknowledges financial support from the Finnish Post Doc Pool and the Jenny and Antti Wihuri Foundation as well as the Jane and Aatos Erkko Foundation for the funding for the LACOR project.

\bibliographystyle{unsrt}
\bibliography{dispersion_bib}

\begin{thebibliography}{1}

\bibitem{KooGor-JPCL-19}
Derk~P. Kooi and Paola Gori-Giorgi.
\newblock A variational approach to london dispersion interactions without
  density distortion.
\newblock {\em The Journal of Physical Chemistry Letters}, 10(7):1537--1541,
  March 2019.

\bibitem{KooWecGor-JCTC-21}
Derk~P. Kooi, Timo Weckman, and Paola Gori-Giorgi.
\newblock Dispersion without many-body density distortion: Assessment on atoms
  and small molecules.
\newblock {\em Journal of Chemical Theory and Computation}, 17(4):2283--2293,
  2021.

\bibitem{BecRou-PRA-89}
A.~D. Becke and M.~R. Roussel.
\newblock Exchange holes in inhomogeneous systems: A coordinate-space model.
\newblock {\em Physical Review A}, 39(8):3761--3767, apr 1989.

\bibitem{KooGor-FD-20}
Derk~Pieter Kooi and Paola Gori-Giorgi.
\newblock London dispersion forces without density distortion: a path to first
  principles inclusion in density functional theory.
\newblock {\em Faraday Discuss.}, 224:145--165, 2020.

\bibitem{ProGanKon-CPL-08}
Emil Proynov, Zhenting Gan, and Jing Kong.
\newblock Analytical representation of the becke--roussel exchange functional.
\newblock {\em Chemical physics letters}, 455(1-3):103--109, 2008.

\bibitem{GorAngSav-CJC-09}
Paola Gori-Giorgi, J{\'{a}}nos~G. {\'{A}}ngy{\'{a}}n, and Andreas Savin.
\newblock Charge density reconstitution from approximate exchange-correlation
  holes.
\newblock {\em Canadian Journal of Chemistry}, 87(10):1444--1450, oct 2009.

\bibitem{JiaMitCheBro-ATDNDT-15}
Jun Jiang, James Mitroy, Yongjun Cheng, and MWJ Bromley.
\newblock Effective oscillator strength distributions of spherically symmetric
  atoms for calculating polarizabilities and long-range atom--atom
  interactions.
\newblock {\em Atomic Data and Nuclear Data Tables}, 101:158--186, 2015.

\bibitem{lda}
Seymour~H Vosko, Leslie Wilk, and Marwan Nusair.
\newblock Accurate spin-dependent electron liquid correlation energies for
  local spin density calculations: a critical analysis.
\newblock {\em Canadian Journal of physics}, 58(8):1200--1211, 1980.

\bibitem{pbe}
John~P Perdew, Kieron Burke, and Matthias Ernzerhof.
\newblock Generalized gradient approximation made simple.
\newblock {\em Physical review letters}, 77(18):3865, 1996.

\end{thebibliography}

\end{document}